\documentclass{article}
\usepackage{spconf,amsmath,graphicx}
\usepackage{hyperref}
\usepackage{float}
\usepackage[export]{adjustbox} 
\usepackage{amsfonts}
\usepackage{booktabs} 
\usepackage{setspace}

\title{HIERARCHICAL NETWORK WITH DECOUPLED KNOWLEDGE DISTILLATION FOR SPEECH EMOTION RECOGNITION}
\name{Ziping Zhao$^{1}$, Huan Wang$^{1}$, Haishuai Wang$^{2}$, Bj{\"o}rn Schuller$^{3,4}$\thanks{The present work is supported by the National Natural Science Foundation of China (No. 62071330).}}
\address{ $^1$College of Computer and Information Engineering, Tianjin Normal University, Tianjin, China\\
         $^2$College of Computer Science, Zhejiang University, China\\
         $^3$Chair of Embedded Intelligence for Health Care and Wellbeing, University of Augsburg, Germany\\
         $^4$GLAM -- Group on Language, Audio, \& Music, Imperial College London, UK\\
}

\begin{document}
\begin{sloppypar} 
\maketitle
\begin{abstract}
The goal of Speech Emotion Recognition (SER) is to enable computers to recognize the emotion category of a given utterance in the same way that humans do. The accuracy of SER is strongly dependent on the validity of the utterance-level representation obtained by the model. Nevertheless, the ``dark knowledge" carried by non-target classes is always ignored by previous studies. In this paper, we propose a hierarchical network, called DKDFMH, which employs decoupled knowledge distillation in a deep convolutional neural network with a fused multi-head attention mechanism. Our approach applies logit distillation to obtain higher-level semantic features from different scales of attention sets and delve into the knowledge carried by non-target classes, thus guiding the model to focus more on the differences between sentiment features. To validate the effectiveness of our model, we conducted experiments on the Interactive Emotional Dyadic Motion Capture (IEMOCAP) dataset. We achieved competitive performance, with 79.1\,\% weighted accuracy (WA) and 77.1\,\% unweighted accuracy (UA). To the best of our knowledge, this is the first time since 2015 that logit distillation has been returned to state-of-the-art status.
\end{abstract}
\vspace{-0.2cm}
\section{Introduction}
\label{sec:intro}
As speech is one of the most natural and direct ways for humans to express emotions, speech emotion recognition (SER) is widely used in fields including online education~\cite{onlinelearning}, psychological healthcare~\cite{depression}, mobile services~\cite{mobile}, etc. The goal of speech emotion recognition is to make the computer extract the emotion that the speaker is expressing from a given utterance. With the development of deep learning, algorithms such as convolutional neural network (CNN), recurrent neural network (RNN), and long short-term memory (LSTM) have been applied to SER tasks~\cite{deepcnn,RNNLSTM,multicnn} to extract abstract features and learn correlations between frames. However, there are still substantial challenges associated with extracting high-level features and precisely classifying emotions. We suggest that logit distillation enables the model to learn knowledge of speech emotion from target to non-target classes, and subsequently acquire higher-level semantic features. However, the cross-entropy in each of these methods focuses only on the target class and ignores the non-target classes.

To address these challenges, we propose a model based on decoupled knowledge distillation with a fused multi-head attention mechanism, namely DKDFMH. The fused multi-head attention mechanism fuses multiple heads in a feature point to facilitate better characterization of the relationship between features. Our motivation for proposing decoupled knowledge distillation for SER is as follows: transferring ``dark knowledge" to students via the teacher's soft labels, and increasing logit similarity between the student and the teacher, which enhances the discriminability of various types of sentiment features.

The main contributions of this paper can be summarized as follows:

1) To the best of our knowledge, this is the first time that decoupled knowledge distillation has been applied to SER.

2) We use decoupled knowledge distillation to overcome the limitations of classical knowledge distillation, utilizing ``dark knowledge" to reduce emotion misclassification and improve accuracy by about 2.9\,\% relative to the current state-of-the-art methods on the IEMOCAP dataset.

3) By decoupling knowledge distillation with a fused multi-head attention mechanism, our model can achieve a weighted accuracy (WA) of 79.1\,\% and unweighted accuracy (UA) of 77.1\,\% on the IEMOCAP dataset.

\vspace{-0.2cm}
\section{RELATED WORK}
\label{sec:related work}
As the field of deep learning has continued to develop, the process of speech emotion recognition has advanced. Acoustic features have progressed from the initial hand-crafted features to the point that specific levels of features can be extracted today. In 2016, Lim et al.~\cite{CNNRNN} proposed a method for extracting audio features by combining CNN with RNN, which yielded results with higher accuracy than could be achieved by traditional manual classification methods. 

Subsequently, the attention mechanism and transformer facilitated significant development in various fields. Aiming to focus more on emotion-related information and reduce the influence of irrelevant external factors, Chen et al.~\cite{3d} proposed a 3-D ACRNN model that combines CNN, Bi-LSTM, and attention mechanisms. Head Fusion was proposed in~\cite{2021HF} by fusing multi-attention heads in the same attention map. In the field of SER, ~\cite{2019CTC+attention, 2022Co-att, 2021CNN-ELM} have shown that the attention mechanism performs well on several datasets, highlighting its effectiveness for sentiment classification.

To learn long-term dependencies in speech signals, Zhao et al.~\cite{selfattention} introduce a self-attention-based knowledge transfer network, in which teacher models learn from speech recognition in order to transfer attention to speech emotion recognition. In previous work, knowledge distillation has been applied to image
classification~\cite{imagedistillation}
and speech recognition~\cite{speechdistillation}, among other fields. In conventional knowledge distillation models, a larger teacher model guides a smaller student model through training by minimizing the loss between the teacher and student models. By contrast, in our approach, the teacher and student models are of the same size, and the weights of the loss values between them can be freely allocated.
\vspace{-0.3cm}
\section{METHODOLOGY}
\subsection{Knowledge Distillation}
In 2015, Hinton et al.~\cite{2015KD} proposed the concept of Knowledge Distillation (KD) to transfer knowledge by minimizing the Kullback–Leibler divergence between the predicted logit of the teacher and student models. Knowledge distillation was subsequently validated on tasks such as speech recognition and image recognition. KD constructively introduces the distillation temperature $T$ into the softmax equation, along with the classification probability $Q=[q_1,q_2,\dots,q_i,\dots,q_C]\in \mathbb{R}^{1\times C}$; the probability of the i-th class is written as $q_i$, the number of classes is written as $C$, the logit of the i-th class is written as $z_i$, such that $q_i$ is calculated as follows:

\begin{equation}
    q_i=\frac{exp(z_i/T)}{\sum_{j=1}^Cexp(z_i/T)} 
\end{equation}
typically, $T$ is set to 1, with one-hot classification used for the label. However, this form of labeling makes the training of the neural network too absolute, meaning that information contained in the incorrect classes is lost. In contrast, when the temperature of $T$ is set to a value greater than 1, the classes whose probability was previously suppressed to 0 will also have a small probability proportional to the distillation temperature. However, this method still limits the potential of logit distillation, as will be discussed in section 3.3.

\begin{figure}[ht]
	\centering
	\includegraphics[width=3.3in, keepaspectratio]{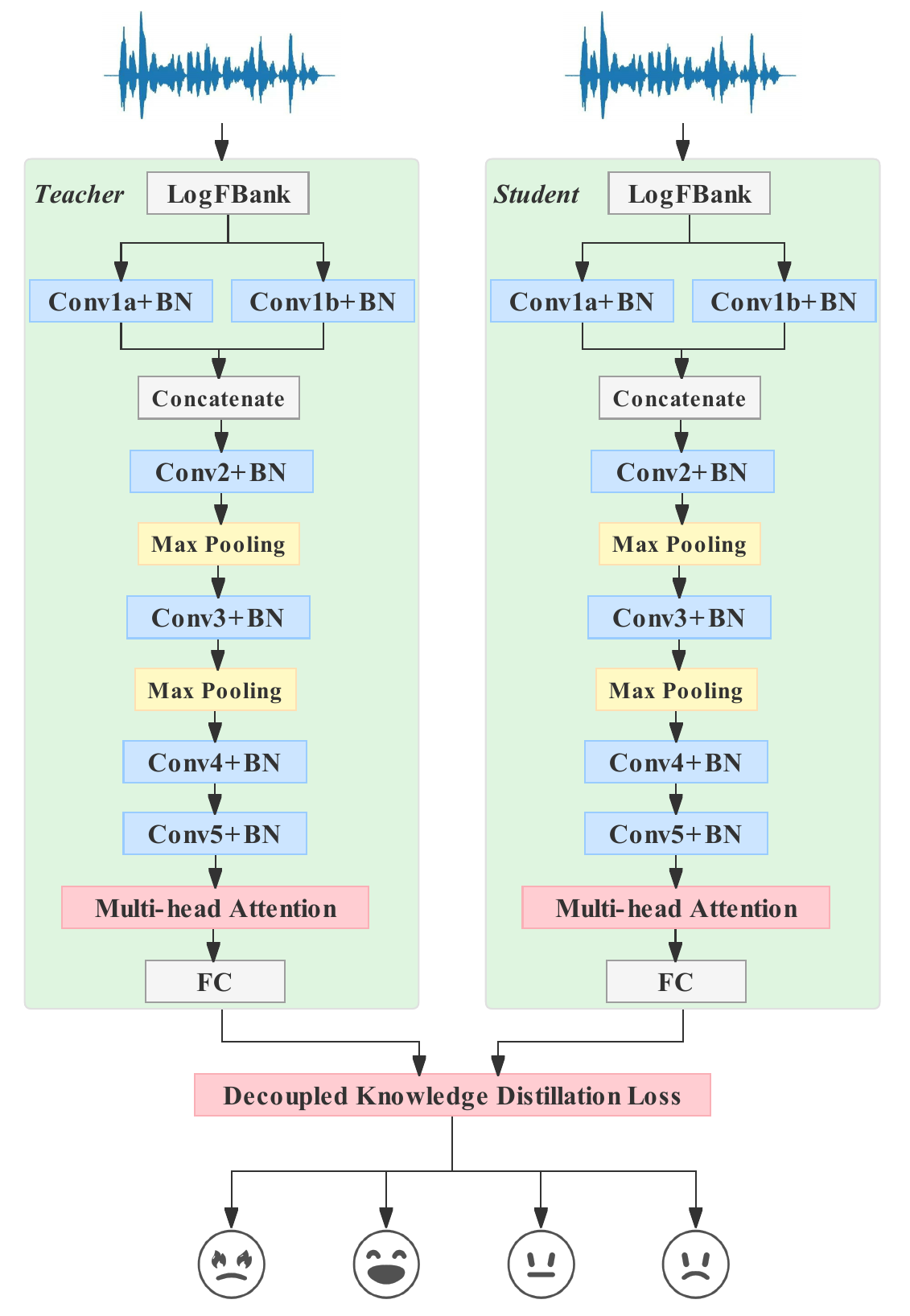}
	\vspace{-0.1cm}
	\caption{Illustration of our DKDFMH model.}
	\vspace{-0.2cm}
	\label{fig:model}
\end{figure}

\vspace{-0.1cm}
\subsection{The teacher and student models}
In cases where the dataset is not too large, and as suggested in the article published by Ji et al.~\cite{ji2021AFD} in 2021, better results can be obtained when the same model is used for both the teacher and student networks. Therefore, we opted to follow this approach. As shown in Fig.\ref{fig:model}, the proposed model is a convolutional neural network with a fused multi-head attention mechanism, which contains a total of five convolutional layers and one attention layer. 

The module input is the log filter bank coefficients (logFBank) spectrum extracted by the Python speech features audio processing library. In the first layer, two parallel convolution layers with kernel sizes of (10,2) and (2,8) are used to extract textures from the temporal and spectral axes, respectively. Each subsequent convolutional layer is followed by batch normalization, with convolutional layers 2 and 3 followed by a max pooling layer with a kernel size of 2 to scale down the data size. After four convolutional layers, we get an 80-channel output, which is fed into the multi-head attention layer to obtain an attention map that maps several different feature attention points, and finally to the fully connected layer for classification.

\vspace{-0.3cm}
\subsection{Decoupled Knowledge Distillation }
Decoupled Knowledge Distillation (DKD), first proposed by Zhao et al.~\cite{2022DKD} in 2022, is the state-of-the-art logit distillation approach. In this paper, we extend decoupled knowledge distillation to the SER task. DKD decouples the logits output by KD into two parts, as in Eq.\ref{equation:reformkd}, using the binary probability $\boldsymbol{b}$ and the probability $\widetilde{\boldsymbol p}$ between non-target classes. $T$ and $S$ represent teachers and students, respectively. The loss function uses Kullback–Leibler (KL) divergence~\cite{kl}, which is defined as follows:
\begin{equation}
	\label{equation:reformkd}
	\setlength{\abovedisplayskip}{6pt}
	\setlength{\belowdisplayskip}{5pt}
	{\rm KD} ={\rm KL}(\boldsymbol b^T||\boldsymbol b^S)  + (1 - p_t^T) {\rm KL}(\widetilde{\boldsymbol p}^T||\widetilde{\boldsymbol p}^S)
\end{equation}
where the knowledge distillation loss is reformulated into the target and non-target classes. ${\rm KL}(\boldsymbol b^T||\boldsymbol b^S)$ represents the similarity of the binary probabilities of the teacher and student for the target class in logit distillation, namely TCKD. ${\rm KL}(\widetilde{\boldsymbol p}^T||\widetilde{\boldsymbol p}^S)$ represents the similarity of the teacher’s and the student’s probabilities for the non-target class, namely NCKD. From this, Eq.\ref{equation:reformkd} can be rewritten as follows:
\begin{equation}
	\label{equation:kd}
	\setlength{\abovedisplayskip}{5pt}
	\setlength{\belowdisplayskip}{5pt}
	{\rm KD} ={\rm TCKD}  + (1 - p_t^T) {\rm NCKD}
\end{equation}
where $p_t^T$ represents the prediction confidence of the teacher network. TCKD is Target Class Knowledge Distillation, while NCKD is Non-Target Class Knowledge Distillation, which transfers ``dark knowledge" via non-target logits.

\begin{figure}[t]
	\centering
	\includegraphics[width=2.8in, keepaspectratio]{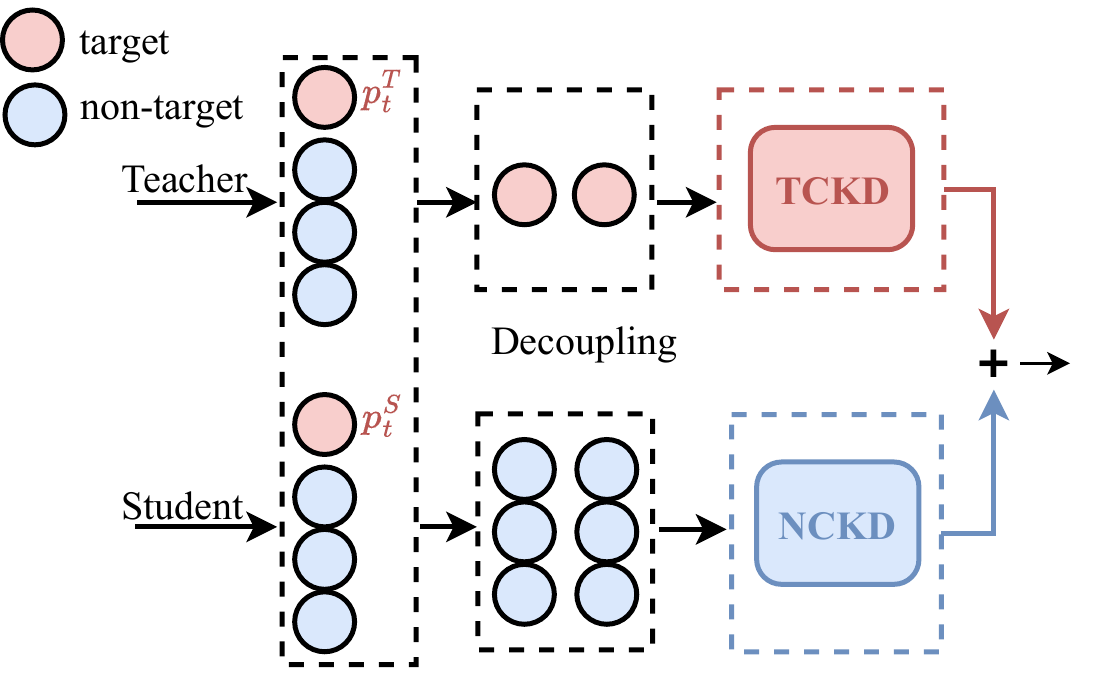}
	\vspace{-0.2cm}
	\caption{Illustration of Decoupled Knowledge Distillation.}
	\label{fig:dkd}
	\vspace{-0.2cm}
\end{figure}

However, as can be seen from Eq.\ref{equation:kd}, NCKD and $p_t^T$ are coupled, and the contribution of NCKD to knowledge distillation is suppressed. To solve the above problem, we use two hyperparameters, $\alpha$ and $\beta$, which are capable of independently adjusting TCKD and NCKD so as to maximize the instruction they provide to the student network. The DKD loss formula is as follows:
\begin{equation}
	\label{equation:dkd}
	{\rm DKD} = \alpha {\rm TCKD} + \beta {\rm NCKD}
	\setlength{\abovedisplayskip}{5pt} 
	\setlength{\belowdisplayskip}{5pt}
\end{equation}
where $\beta$ replaces $(1 - p_t^T)$, which would be suppressed. As shown in Fig.\ref{fig:dkd}, DKD provides weights that enable free balance, allowing the model to learn the most appropriate loss.

In the SER context, some emotions have characteristics that closely resemble each other, making it quite challenging for conventional models to distinguish between these emotions. For example, happiness and anger are often misclassified in the predictions. Notably, however, our model can acquire higher-level semantic features via DKD loss and thereby guide the teacher model to deliver more ``dark knowledge" to the student model. Moreover, compared to feature distillation, which requires additional computation, storage, and complex structures to align dimensions, logit distillation is also simpler and more efficient.
\vspace{-0.3cm}
\section{EXPERIMENTS}
\label{sec:typestyle}

\vspace{-0.2cm}
\subsection{Dataset}
\label{sec:dataset}
We use the well-benchmarked corpus Interactive Emotional Dyadic Motion Capture (IEMOCAP)~\cite{2008iemocap}, which was collected by the University of Southern California. IEMOCAP contains five sessions, each performed by a pair of subjects (one male and one female) in scripted and improvised scenarios. The dataset contains approximately 12 hours of audiovisual data. The average duration of each voice segment is 4.5 sec. 

In previous experiments~\cite{li2018attention,tarantino2019self}, the accuracy on improvised data was higher than that on scripted data; this may be because the actors delivered more emotionally realistic performances during improvisation. In this paper, we choose to use improvised data, with four types of emotions: angry, happy, neutral, and sad. Due to the imbalanced data distribution and the fact that excitement and happiness data tend to be highly similar in the activation and valence domains, most researchers choose to either replace excitement with happiness or to combine the data for both; our experiments use the former approach.

\vspace{-0.3cm}
\subsection{Evaluation Metrics}
\label{sec:evaluation metrics}
Weighted accuracy (WA) and unweighted accuracy (UA) are employed to validate the predictive performance of our proposed model. UA is calculated as the average accuracy of the emotional categories, while WA is calculated as the accuracy of all samples. These two evaluation methods are both widely utilized in contemporary SER research.

\begin{figure}[h]
	\centering
	\includegraphics[width=217pt, keepaspectratio]{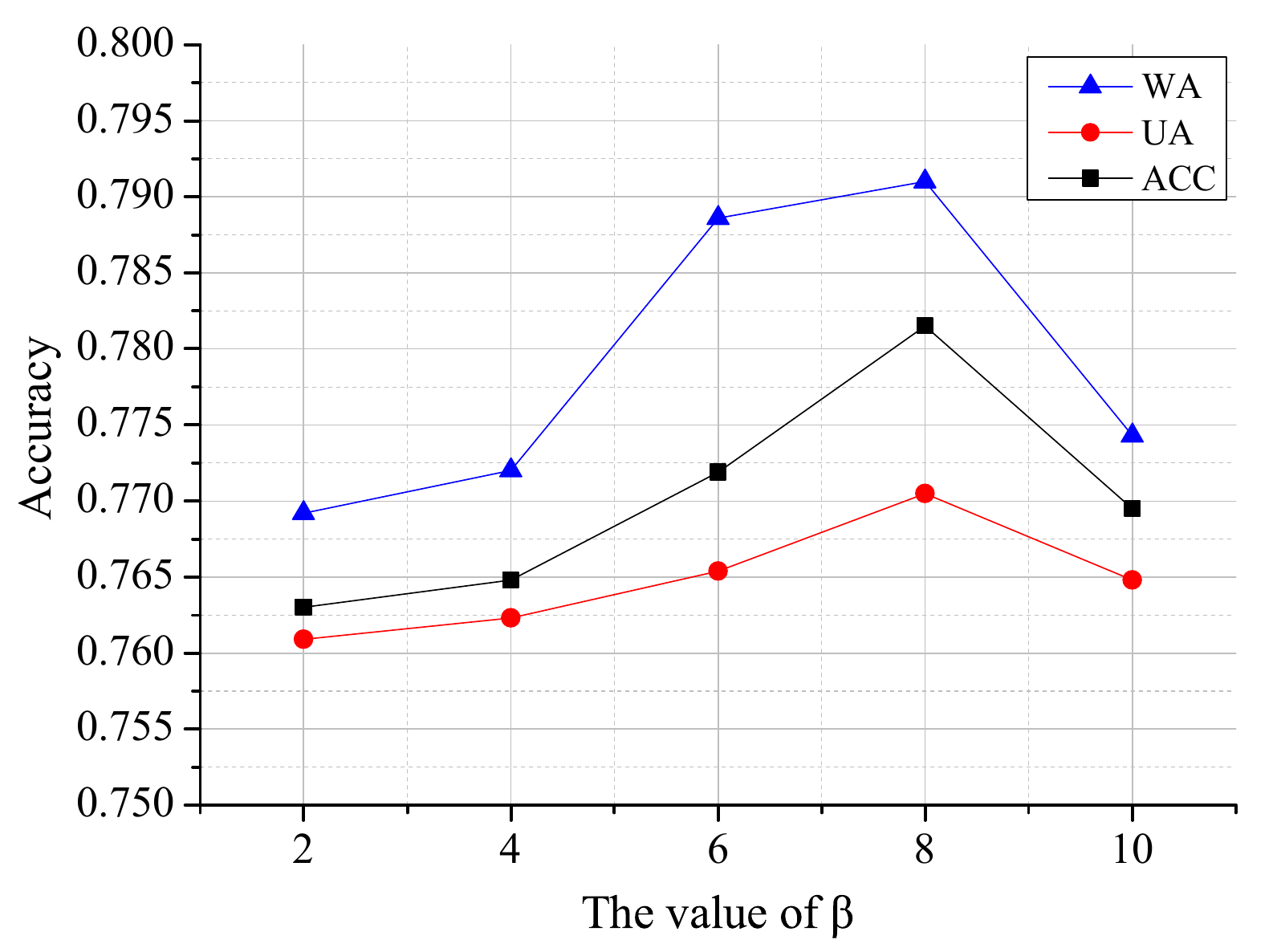}
	\vspace{-0.2cm}
	\caption{Results of modifying $\beta$.}
	\label{fig:value of β}
\end{figure}
\vspace{-0.3cm}
\subsection{Experimental Setup}
\label{sec:experimental setup}
We implement our proposed model in PyTorch and randomly split the dataset into 80\% for training and 20\% for testing. We use LogFBank extracted by 40 filters as the feature input, which yields a 197-dimension feature vector. Compared with MFCC, this approach requires comparatively less computation, and the correlation of each feature is made stronger.

In the feature extraction process, we set the window length to 0.04 sec and the step size to 0.01 sec. Each utterance is divided into 2-sec segments, while there is a 1-sec overlap between each segment in the training set and a 1.6-sec overlap in the testing set. The DKD loss is employed in our experiments; $\alpha$ is set to 1, $\beta$ is set to 8, and the distillation temperature $T$ is set to 4. The optimizer used is Adam, and the system is trained with a batch size of 32 for 50 epochs. The initial learning rate is $1\times10^{-4}$, and the decay rate is $1\times10^{-6}$.

\begin{table}[t]
	\caption{Ablation study on our model(\%).}
	\label{tab:table1}
	\centering
	\begin{tabular}{ccc}
		\toprule
		\textbf{Method}  & \textbf{WA}  & \textbf{UA}   \\
		\midrule
		CNN      & 74.5          & 72.2          \\ 
		CNN+multi-head attention      & 76.2          & 74.8          \\ 
		CNN+multi-head attention+KD  & 76.9  & 76.4    \\ \hline
		TCKD              & 75.3          & 72.6          \\ 
		NCKD              & 77.7          & 76.9          \\ 
		DKDFMH & \textbf{79.1} & \textbf{77.1} \\ 
		\bottomrule
	\end{tabular}
\vspace{-0.3cm}
\end{table}

\vspace{-0.3cm}
\subsection{Experimental Results}
\label{sec:experimental results}
{\bf Ablation study} In this section, we verify the effectiveness of DKDFMH through ablation experiments: the model only using CNN, the model using CNN and multi-head attention, the model with knowledge distillation added, the model with TCKD separately, the model with NCKD separately, and the model with decoupled knowledge distillation. 

As can be seen from Table\ref{tab:table1}, using decoupled knowledge distillation resulted in the highest accuracy. The addition of the knowledge distillation improves the weighted accuracy by nearly 1.0\%, indicating that the incorrect classes also contain information. We use the results of the knowledge distillation model as the baseline, then compare this with the results of the experiments using TCKD and NCKD respectively. 

It is obvious from the experimental results that the use of TCKD as a separate model is detrimental to the distillation effect, reducing the accuracy of the model by 1.7\% to 75.3\%. In contrast, using NCKD separately for distillation is about the same or even better than using knowledge distillation, leading to a 0.7\% to 77.7\% increase in the weighted accuracy. We can accordingly conclude that knowledge between non-target classes is crucial for logit distillation, NCKD is the main reason why logit distillation is effective but suppressed; after decoupled KD, we can freely adjust the coefficients of TCKD and NCKD to produce better distillation performances.
\begin{figure}[t]
	\centering
	\includegraphics[width=0.7\linewidth]{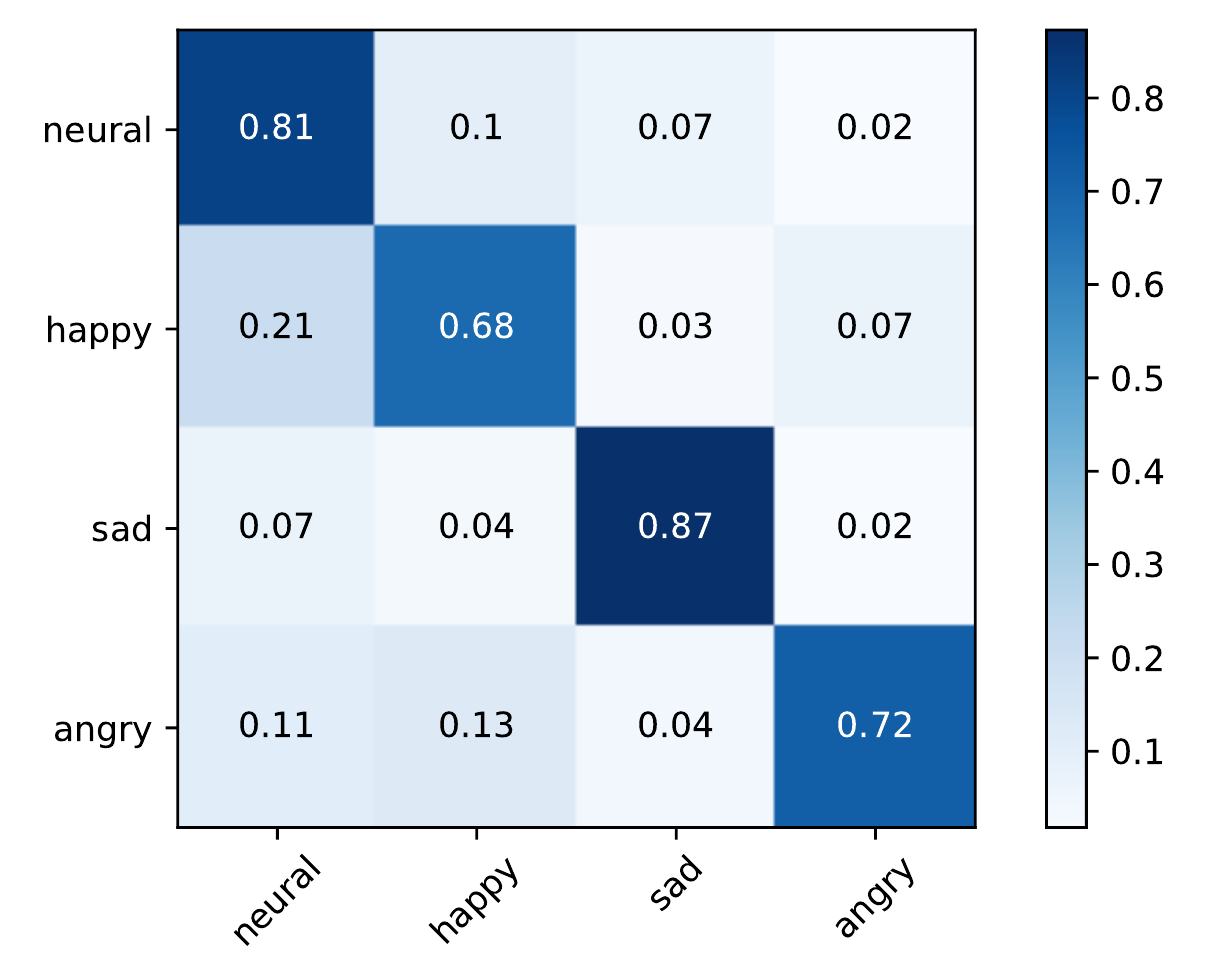}
	\vspace{-0.3cm}
	\caption{Confusion Matrix of IEMOCAP.}
	\vspace{-0.4cm}
	\label{fig:confusion-matrix}
\end{figure}

\noindent
{\bf Effect of $\beta$} Experiments are also conducted to study the effect of the setting of $\beta$ on the SER performance when $\alpha$ = 1. As shown in Fig.\ref{fig:value of β}, accuracy is maximized when $\beta$ = 8.

\noindent
{\bf Comparison with state-of-the-art methods} As shown in Table\ref{tab:table 2}, our proposed model achieves the best performance when compared to the state-of-the-art methods (under the premise of using the same dataset and evaluation metrics). Compared to the Head Fusion employed by Xu et al.~\cite{2021HF}, the application of decoupled knowledge distillation improved the overall model accuracy by 2.9\%. The model proposed by Zou et al.~\cite{2022Co-att} also incorporated a multilayer attention mechanism; DKDFMH outperforms it by 7.5\% in terms of WA and 4.3\% in terms of UA, proving the outstanding generalizability and distinguishability of the features learned through DKDFMH. The advantages of our model for similar emotion recognition can be more intuitively demonstrated by the confusion matrix in Fig.\ref{fig:confusion-matrix}.

\vspace{-0.3cm}
\begin{table}[ht]
\caption{Comparison with state-of-the-art methods (\%).}
\label{tab:table 2}
\centering
\begin{tabular}{ccc}
   \toprule
   \textbf{Method} & \textbf{WA} & \textbf{UA} \\
   \midrule
   CTC+Attention~\cite{2019CTC+attention} & 67.0 & 69.0  \\
  Head Fusion~\cite{2021HF}  & 76.2 & 76.4 \\
  HGFM~\cite{2020HGFM} & 66.6 & 70.5  \\
  DAAE+CNN+Attention~\cite{2021DAAE} & 70.1 & 70.7 \\
  HNSD~\cite{2021HNSD} & 70.5 & 72.5 \\
  CNN-ELM+STC attention~\cite{2021CNN-ELM} & 61.3 & 60.4 \\
  Multi-level Co-att~\cite{2022Co-att} & 71.6 & 72.7 \\ \hline
  DKDFMH & \textbf{79.1} & \textbf{77.1}  \\
   \bottomrule
\end{tabular}
\vspace{-0.3cm}
\end{table}

\vspace{-0.3cm}
\section{conclusion}
\label{sec:conclusion}
In this paper, we apply decoupled knowledge distillation to SER and a convolutional neural network fusing multi-head attention to the teacher and student networks. A WA of 79.1\% and a UA of 77.1\% were obtained on the IEMOCAP dataset, proving the great potential of our proposed model. In the future, we will continue to study the application of knowledge distillation in SER. Moreover, in order to more closely replicate real-world SER scenarios, we will try to add noise to the audio data so as to improve the robustness of the model.



\bibliographystyle{IEEEbib}
\bibliography{refs}   

\end{sloppypar}
\end{document}